\documentclass[prd,superscriptaddress,a4paper,showpacs,showkeys,nofootinbib]{revtex4}
\usepackage{graphicx}
\usepackage{subfigure}
\usepackage{rotating}
\usepackage{epsfig}
\usepackage[lofdepth,lotdepth,caption=false]{subfig}
\usepackage{bm}
\usepackage{amsmath}
\usepackage{dcolumn}
\usepackage{bm}

\usepackage{url}
\usepackage{amsmath,amssymb,graphicx}
\usepackage{amsmath}
\usepackage{amssymb}
\usepackage{mathrsfs}
\usepackage{accents}
\usepackage{epsfig}
\usepackage{enumerate}

\usepackage{mathtools}
\usepackage{mathrsfs}

\usepackage{color}

\newcommand{\rr}{\mbox{\boldmath $r$}}

\newcommand{\rb}{\mbox{\boldmath $b$}}

\begin{document}

\begin{flushright}
MS-TP-22-47
\end{flushright}

\title{Higher twists effects in DIS on nuclei at the EIC and LHeC: \\ A phenomenological analysis}


\author{Yan B. {\sc Bandeira}}
\email{yan.bandeira@ufpel.edu.br}
\affiliation{Institute of Physics and Mathematics, Federal University of Pelotas, \\
  Postal Code 354,  96010-900, Pelotas, RS, Brazil}

\author{Victor P. {\sc Gon\c{c}alves}}
\email{barros@ufpel.edu.br}
\affiliation{Institut f\"ur Theoretische Physik, Westf\"alische Wilhelms-Universit\"at M\"unster,
Wilhelm-Klemm-Straße 9, D-48149 M\"unster, Germany
}
\affiliation{Institute of Modern Physics, Chinese Academy of Sciences,
  Lanzhou 730000, China}
\affiliation{Institute of Physics and Mathematics, Federal University of Pelotas, \\
  Postal Code 354,  96010-900, Pelotas, RS, Brazil}


\begin{abstract}
In this paper we investigate the impact of the higher - twist effects, resummed by the non - linear approaches for the QCD dynamics, on the inclusive observables that will be measured in future electron - ion colliders. We assume a phenomenological model for the dipole - nucleus scattering amplitude which takes into account the non - linear corrections and estimate the contribution of the different twists for $F_2^A(x,Q^2)$, $F_L^A(x,Q^2)$ and $\partial F_2^A /\partial ln Q^2$ considering different values of the Bjorken - $x$ variable, photon virtuality $Q^2$ and atomic number $A$. A comparison with the full predictions is performed and the impact of the distinct twists is estimated. Our results indicate that a future analysis of the logarithmic $Q^2$ slope and longitudinal structure function will allow us to probe the presence of the non - linear effects on the QCD dynamics.
\end{abstract}
\maketitle

\section{Introduction}
The study of the hadron structure at high energies is one of the main goals of the  high energy colliders \cite{AbdulKhalek:2021gbh,LHeC:2020van}. 
From the analyses performed in the $ep$ collider at HERA, we know that 
the proton structure function increases with the decreasing of the Bjorken - $x$ variable,  that 
 a large fraction of the observed events ($\approx 10 \%$) are diffractive and that the total  cross sections   present the property of geometric scaling \cite{Klein:2008di}. In addition, from the studies of  particle production at forward rapidities in  $pA$ collisions at RHIC and LHC, one has that the high $p_T$ hadron yields are strongly suppressed \cite{hdqcd}. All these results can be quite well described by models based on Color Glass Condensate (CGC) formalism \cite{CGC}, which take into account non - linear effects that are inherent to  the QCD dynamics at high energies when the hadron becomes a dense system. The transition line between the linear and non - linear regimes of the QCD dynamics is described by the saturation scale $Q_s$, which is predicted to depend on $x$ and  atomic number $A$. Although several experimental results provide some evidence for the presence of non - linear effects in the QCD dynamics, more definite conclusions are not possible due to the small value of the saturation scale $Q_s$ in the kinematical range of HERA and due to the complexity of hadronic collisions.  As $Q_s$ increases at smaller values of $x$ and larger values of $A$, future electron - nucleus  colliders \cite{AbdulKhalek:2021gbh,LHeC:2020van} are expected to provide the ideal scenario to determine whether parton distributions saturate or not and allow us to disentangle non - linear  from linear physics.

One important characteristic of the non - linear approaches is that they resum higher  twist
contributions \cite{Bartels:1999xt,Bartels:2000hv,Bartels:2009tu,Boussarie:2021ybe}, i.e. they contain information of all orders in $1/Q^2$ in the perturbative expansion\footnote{In the Operator Product Expansion (OPE),  the scattering amplitudes are expanded into a series of contributions  $\sigma = \sum_{\tau} \sigma_{\tau} (Q^2)$, with $Q^2$ being the characteristic hard scale,  $\sigma_{\tau} \propto 1/Q^{\tau}$ and $\tau = 2, 4, \,...$ being the twist \cite{Ellis:1982cd}.}, which become important for small values of the hard scale $Q^2$. In contrast, the linear approaches based on the collinear factorization and DGLAP evolution \cite{dglap} are { usually } formulated at leading twist ($\tau = 2$).  The difference between the linear and non - linear predictions is expected to increase at low values of $Q^2$ and smaller values of $x$. The study performed in Ref. \cite{Motyka:2017xgk}, using the combined HERA data for the inclusive cross sections, has indicated that the higher twist contributions to the proton structure are comparable to the leading twist contributions at low $Q^2 \approx 2$ GeV$^2$ and $x \approx 10^{-5}$ (See also Ref.  \cite{Abt:2016vjh}). 
Moreover, the studies performed in Refs. \cite{Goncalves:2004yc,Illarionov:2004nw,Motyka:2012ty,Badelek:2022cgr,Maktoubian:2019ppi,Goharipour:2020gsw} have pointed out that the impact of the higher twist corrections is still larger on the longitudinal structure function and in diffractive observables. As in the saturation approaches the magnitude of the higher twist corrections is determined by the saturation scale, it is natural to expect a larger impact on the observables that will be measured in $eA$ collisions for the energies of the proposed Electron - Ion Collider (EIC) at BNL \cite{AbdulKhalek:2021gbh} and Large Hadron Electron Collider (LHeC) at CERN \cite{LHeC:2020van}.
Such an aspect was already pointed out in Ref. \cite{Gotsman:2000fy}, which has presented a first estimate of these effects for DIS on nuclei. Our goal in this paper is twofold. First, we will update the analysis performed in Ref. \cite{Gotsman:2000fy}, presenting a systematic study of the higher twist corrections on  $F_2^A(x,Q^2)$ and $F_L^A(x,Q^2)$,  considering the kinematical range of the future colliders and distinct values of the photon virtuality and atomic number. A comparison with the proton case will also be presented. Second, we will estimate, for the first time, the impact of these corrections on the logarithmic $Q^2$ slope of the nuclear structure function, $\partial F_2^A /\partial ln Q^2$. 
As it is well known, at leading twist, such a quantity is a direct probe of the  gluon distribution \cite{Prytz:1993vr,GayDucati:1996ry}. Moreover, as demonstrated in Refs. \cite{GayDucati:1999xg,Goncalves:2000ex}, the slope is sensitive to the saturation effects and a direct probe of the saturation scale. 

\begin{figure}[t]
    \centering
    \includegraphics[scale=0.35]{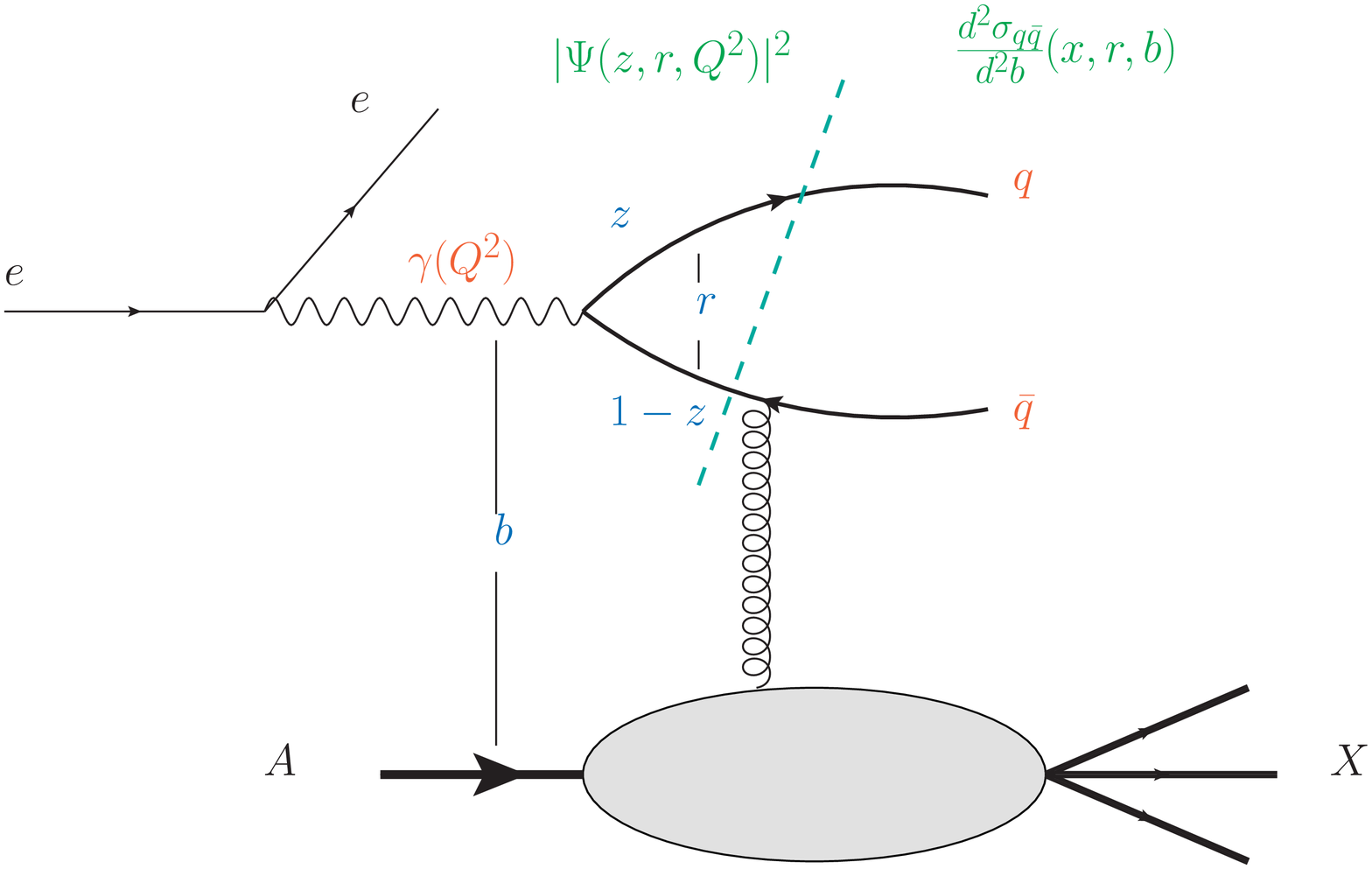}
    \caption{Representation of the deep inelastic scattering on nuclei in the dipole formalism.}
    \label{Fig:diagram}
\end{figure}

 This paper is organized as follows. In the next Section, we will present a brief review of the dipole formalism for the calculation of inclusive observables in $eA$ collisions. The phenomenological model for the dipole - nucleus scattering used in our study will be discussed  and the twist expansion  of the longitudinal and transverse structure functions will be presented.  In Section \ref{Sec:res} we will present our predictions for the $x$ - dependence of 
$F_2^A(x,Q^2)$, $F_L^A(x,Q^2)$ and $\partial F_2^A /\partial ln Q^2$ considering different values of the atomic number and photon virtuality. In particular, the sensitivity of the maximum in the slope on the saturation effects will be demonstrated. Moreover, the predictions of the different twists for the observables are presented considering distinct values of $A$ and $Q^2$.   Finally, in Section \ref{sec:conc} we will summarize our main results and conclusions.

\section{Formalism}
\label{sec:form}
In what follows we will present a brief review of the dipole formalism \cite{nik} for the description of the DIS on nuclei. Such process is characterized by a large electron energy loss $\nu$ (in the target rest frame) and an invariant momentum transfer $q^2 \equiv - Q^2$  between the incoming and outgoing electrons, such that $x = Q^2 / 2 m_N \nu$ is fixed.
In the Breit frame, the photon probes the partonic structure of the nucleus, with the quark and gluon contents increasing for smaller values of $x$. One has that the high partonic densities present in this regime implies the overlap  in the longitudinal direction of the parton clouds originated from different bound nucleons. As a consequence,  low $x$ partons from different nucleons overlap spatially creating much larger parton densities than in the free nucleon case, implying the amplification of the non - linear effects in the QCD dynamics \cite{Gribov:1983ivg}. On the other hand, in the dipole frame, the $eA$ scattering can be viewed in terms of the photon emission by the electron $|e\rangle \rightarrow |e \gamma \rangle$, the splitting of the photon in a color dipole $|e \gamma \rangle \rightarrow |e q \bar{q} \rangle$ and the propagation of the dipole in the gluonic field of the nucleus. At high QCD field strengths, the multiple scatterings of the dipole with the nucleus become relevant and must be taken into account. In this frame,  the nuclear  structure functions can be factorized  as follows (See Fig. \ref{Fig:diagram})
\begin{eqnarray}
    F_{L,T} (x,Q^2) & = & \frac{Q^2}{4\pi^2\alpha_{\text{em}}} \sigma_{L,T}^{\gamma^*p}(x,Q^2) \nonumber \\
& = & \frac{Q^2}{4\pi^2\alpha_{\text{em}}} \int dz \int d^2\rr \int d^2\rb \,\, |\Psi_{L,T}(z,\rr,Q^2)|^2 \, \frac{d^2 \sigma}{d^2\rb} (x,\rr,\rb) \,\,,
\label{Eq:fs}
\end{eqnarray}
where $\rr$ denotes the transverse size of the dipole, $z$ is the longitudinal momentum fraction carried by the quark, $\rb$ is the transverse distance from the center of the nucleus to the center of mass of the $q \bar{q}$  dipole and  $\Psi_{T,L}$ are the wave functions of the  photon corresponding to their transverse  or longitudinal polarizations. Explicit expressions for $\Psi_{T,L}$ are given, e.g., in Ref. \cite{nik}.    Furthermore,  $d^2 \sigma/d^2\rb$ describes the dipole - nucleus interactions and its description is directly associated with the QCD dynamics \cite{hdqcd}.
In recent years, several groups have proposed different phenomenological approaches to describe this quantity, which are  based on the Color Glass Condensate (CGC) formalism \cite{CGC} and successfully describe a large set of observables in $ep$, $pp$, $pA$, and $AA$ collisions. However, the description of the impact parameter dependence of the dipole - nucleus cross section and the impact of next - to - leading order corrections on the observables are still themes of intense study (For a recent discussion see, e.g. Refs. \cite{Bendova:2022xhw,Bendova:2020hkp}). As our goal in this exploratory study is to investigate the impact on the higher twist corrections on the inclusive observables that will be analyzed in future $eA$ colliders, in what follows we will assume that the impact parameter dependence of $d^2 \sigma/d^2\rb$, which is mainly associated to non - perturbative physics, can be factorized and its dependence on the saturation scale is given as in the phenomenological model proposed in Ref. \cite{GBW}. Therefore, we will assume that 
\begin{eqnarray}
\frac{d^2\sigma}{d^2\rb} (x,\rr,\rb) = 2 \, (1 - e^{ - r^2 Q_{s,A}^2(x) /4}) S(\rb) \,\,,
\label{Eq:dA}
\end{eqnarray}
where $r \equiv |\rr|$, $Q_{s,A}(x)$ is the saturation scale for the nucleus and $S(\rb)$ is the profile function in impact parameter space, { which is usually described by a Wood - Saxon distribution}. Theoretically, we expect that $Q_{s,A}^2 = A^{\alpha} \times Q_{s,p}^2$, with $\alpha \approx 1/3$, and that $\int d^2\rb S(\rb) \propto A^{2/3}$.  In order to simplify our analysis and to make contact with previous studies, in what follows we will assume that $\alpha = 1/3$, $Q_{s,p}^2 = Q_0^2 (x_0/x)^{\lambda}$ and $\int d^2\rb S(\rb) = \sigma_0^A/2 = A^{2/3} \sigma_0/2$, with the values for  $\sigma_0$, $Q_0$, $x_0$ and $\lambda$ fixed by the fit to the $ep$ HERA data performed in Ref. \cite{GBW}. Such assumptions imply that the results derived in Ref. \cite{Bartels:2000hv} can be directly generalized for a nuclear target.

Following Ref. \cite{Bartels:2000hv}, we can employ a Mellin transform of Eq. (\ref{Eq:fs}) and  perform an expansion in powers of $\xi \equiv Q_{s,A}^2 /Q^2$. Such procedure implies that the leading twist ($\tau = 2$) contributions for the transverse and longitudinal structure functions will be given by \cite{Bartels:2000hv}
\begin{eqnarray}
    F_T^{\tau = 2}(x,Q^2) & = & \frac{Q^2}{4\pi^2\alpha_{\text{em}}} \sigma_0^A  \sum_f e^2_f \frac{\alpha_{\text{em}}}{\pi} \left[\frac{7}{6}\xi - \psi(2)\xi + \xi\ln\left(\frac{1}{\xi}\right)\right] \,\,, \\
    F_L^{\tau = 2}(x,Q^2) & = &  \frac{Q^2}{4\pi^2\alpha_{\text{em}}} \sigma_0^A \sum_f e^2_f \frac{\alpha_{\text{em}}}{\pi} \xi \,\,,
\label{Eq:tw2}
\end{eqnarray}
{ where $e_f$ is the quark's electric charge in units of the electron charge and $\psi(x)$ is the digamma function}.
Moreover, one also has the following expressions for the higher twists:
\begin{itemize}
\item Twist - 4:
\begin{eqnarray}
    F_T^{\tau = 4}(x,Q^2) & = & \frac{Q^2}{4\pi^2\alpha_{\text{em}}} \sigma_0^A  \sum_f e^2_f \frac{\alpha_{\text{em}}}{\pi}\frac{6}{10}\xi^2 \,\,, \\
    F_L^{\tau = 4}(x,Q^2) & = &  \frac{Q^2}{4\pi^2\alpha_{\text{em}}} \sigma_0^A  \sum_f e^2_f \frac{\alpha_{\text{em}}}{\pi} \left[- \frac{94}{75}\xi^2 + \frac{4}{5}\psi(3)\xi^2 - \frac{4}{5}\xi^2 \ln\left(\frac{1}{\xi}\right)\right] \,\,.
\label{Eq:tw4}
\end{eqnarray}
\item Twist - 6:
\begin{eqnarray}
    F_T^{\tau = 6}(x,Q^2) & = & \frac{Q^2}{4\pi^2\alpha_{\text{em}}} \sigma_0^A  \sum_f e^2_f \frac{\alpha_{\text{em}}}{\pi}\left[\frac{43}{1225}\xi^3 - \frac{12}{35}\psi(4)\xi^3 + \frac{12}{35}\xi^3\ln\left(\frac{1}{\xi}\right) \right] \,\,, \\
    F_L^{\tau = 6}(x,Q^2) & = &  \frac{Q^2}{4\pi^2\alpha_{\text{em}}} \sigma_0^A \sum_f e^2_f \frac{\alpha_{\text{em}}}{\pi} \left[ \frac{654}{1225}\xi^3 - \frac{36}{35}\psi(4)\xi^3 + \frac{36}{35}\xi^3\ln\left(\frac{1}{\xi}\right)\right] \,\,.       
\label{Eq:tw6}
\end{eqnarray}
\item Twist - 8:
\begin{eqnarray}
    F_T^{\tau = 8}(x,Q^2) & = & \frac{Q^2}{4\pi^2\alpha_{\text{em}}} \sigma_0^A  \sum_f e^2_f \frac{\alpha_{\text{em}}}{\pi}\left[-\frac{262}{11025}\xi^4 + \frac{4}{35}\psi(5)\xi^4 - \frac{4}{35}\xi^4 \ln\left(\frac{1}{\xi}\right) \right] \,\,,  \\
    F_L^{\tau = 8}(x,Q^2) & = &  \frac{Q^2}{4\pi^2\alpha_{\text{em}}} \sigma_0^A \sum_f e^2_f \frac{\alpha_{\text{em}}}{\pi}\left[-\frac{1636}{18375}\xi^4 + \frac{48}{175}\psi(5)\xi^4 - \frac{48}{175}\xi^4 \ln\left(\frac{1}{\xi}\right)
\right]\,\,.  
\label{Eq:tw8}
\end{eqnarray}
\end{itemize}
Such expressions can be used to derive the twist expansion of $F_2^A(x,Q^2) = F_T^A(x,Q^2)+ F_L^A(x,Q^2)$ and of the logarithmic $Q^2$ slope of $F_2^A(x,Q^2)$. It is important to emphasize that the above expressions are valid for $\xi \lesssim 1$, i.e. in the linear regime, where $Q^2 \gg Q_{s,A}^2$, and near to the transition line between the linear and non - linear regimes, which is characterized by $Q^2 \approx Q_{s,A}^2$. As the saturation scale increases with the atomic number and with the decreasing of the Bjorken - $x$ variable, one should be careful in the use of these expressions for a large nuclei, low $Q^2$ and $x \rightarrow 0$. 

\begin{figure}[t]
\begin{center}
{\includegraphics[width=.6\textwidth]{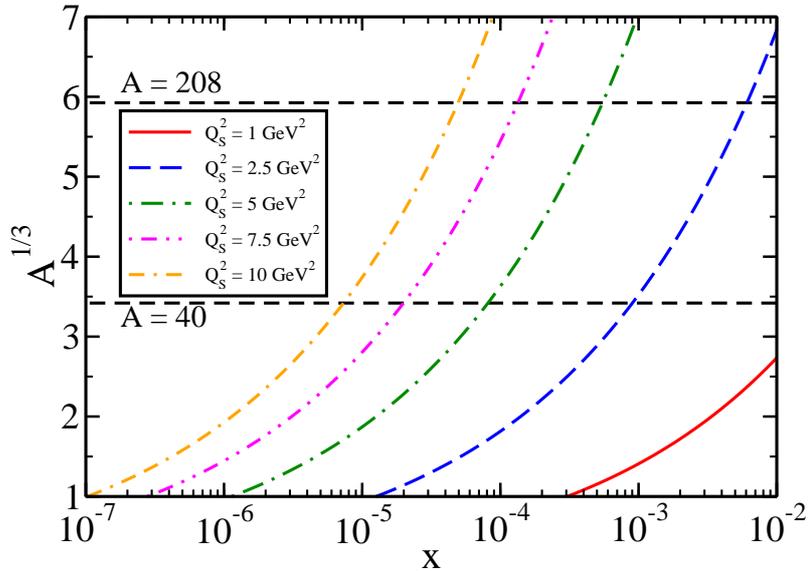}}
\caption{Nuclear saturation scale as a function of the Bjorken $x$ and nuclear mass number $A$.}
\label{qsa}
\end{center}
\end{figure}

{ In order to clarify the region of validity of our study  let's analyze the $x$ and nuclear dependence of the saturation scale. In our study we will assume that the saturation scale of the nucleus is enhanced with respect to the nucleon one by the {\it oomph} factor $A^{\frac{1}{3}}$, with $Q_{s,A}^2 = A^{\frac{1}{3}} \times Q_0^2 \, (\frac{x_0}{x})^{\lambda}$, which implies that the nuclei are an efficient amplifier of the non-linear effects. In Fig. \ref{qsa} we present the theoretical expectations for the saturation scale as a function of $x$ and $A$, considering the  parameters  $Q_0^2 = 1.0$ GeV$^2$, $x_0 = 3.04 \times 10^{-4}$ and $\lambda = 0.288$ as in Ref. \cite{GBW}. We can observe that, while in the proton case we need very 
small values of $x$ to obtain large values of $Q_s^2$, in the nuclear case a similar value 
can be obtained for values of $x$ approximately two orders of magnitude greater. Therefore, the parton density that will be accessed at the EIC is equivalent to that obtained in $ep$ colllisions at energies 
that are at least one order of magnitude higher than at HERA. Moreover, these results can be used to estimate the region of validity of the twist expansion considered in this paper. The results are valid for $\xi \lesssim 1$,  where $Q^2 \geq Q_{s,A}^2$. As a consequence, from Fig. \ref{qsa} one has e.g. that for  $Q_s^2 = 2.5$ GeV$^2$, the calculations will be valid for $x \gtrsim 10^{-5} / 7 \times 10^{-4} / 5 \times 10^{-3}$ when $A = 1 / 40 /208$.   Such aspects will be considered in the analysis of the results presented in the next Section.}

\begin{figure}[t]
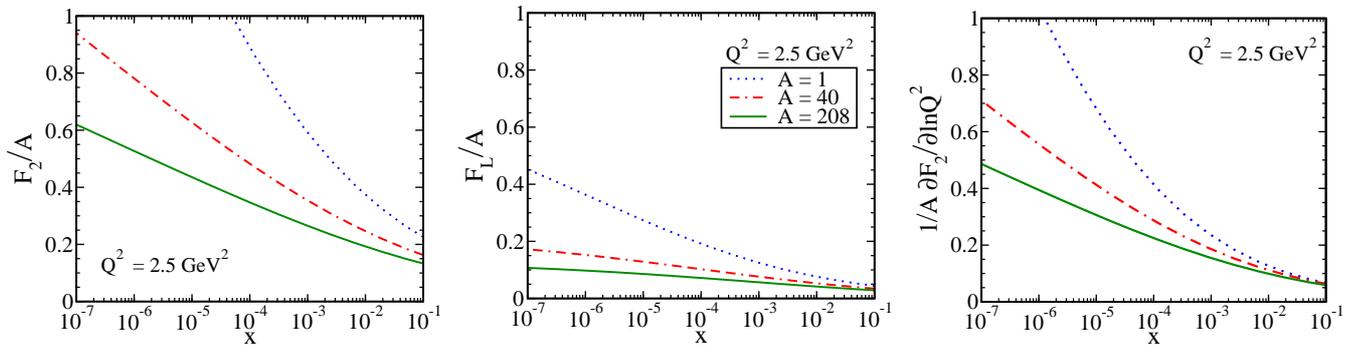

\centering
    \includegraphics[width=.325\textwidth]{GBW_F2_Q2_2_5.eps}
    \includegraphics[width=.325\textwidth]{GBW_FL_Q2_2_5.eps}
    \includegraphics[width=.325\textwidth]{diff_GBW_F2_Q2_2_5.eps}
\caption{Predictions for $F_2^A(x,Q^2)$, $F_L^A(x,Q^2)$ and $\partial F_2^A /\partial ln Q^2$, normalized by $A$, considering different values of the atomic number and $Q^2 = 2.5$ GeV$^2$. }
\label{Fig:Full_A}
\end{figure}

\begin{figure}[t]
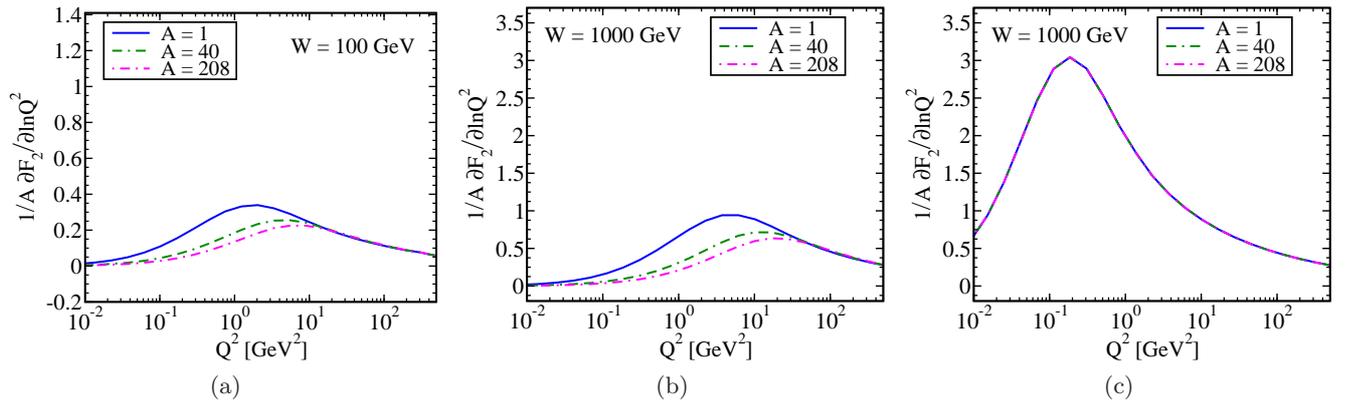

    \centering
\begin{tabular}{ccc}
   \includegraphics[width=0.32\textwidth]{Diff_GBW_W_100_Q2.eps}  &
    \includegraphics[width=0.32\textwidth]{Diff_GBW_W_1000_to_comparison_Q2.eps} &
        \includegraphics[width=0.32\textwidth]{Diff_GBW_W_1000_linear_Q2.eps} \\
        (a) & (b) & (c)

\end{tabular} 
    \caption{Predictions for $\partial F_2^A /\partial ln Q^2$, normalized by $A$, considering different nuclei and (a) $W = 100$ GeV and (b) $W = 1000$ GeV. The predictions for $W = 1000$ GeV, derived disregarding the non - linear effects, are presented in panel (c). }
    \label{Fig:slope1}
\end{figure}

\begin{figure}[t]
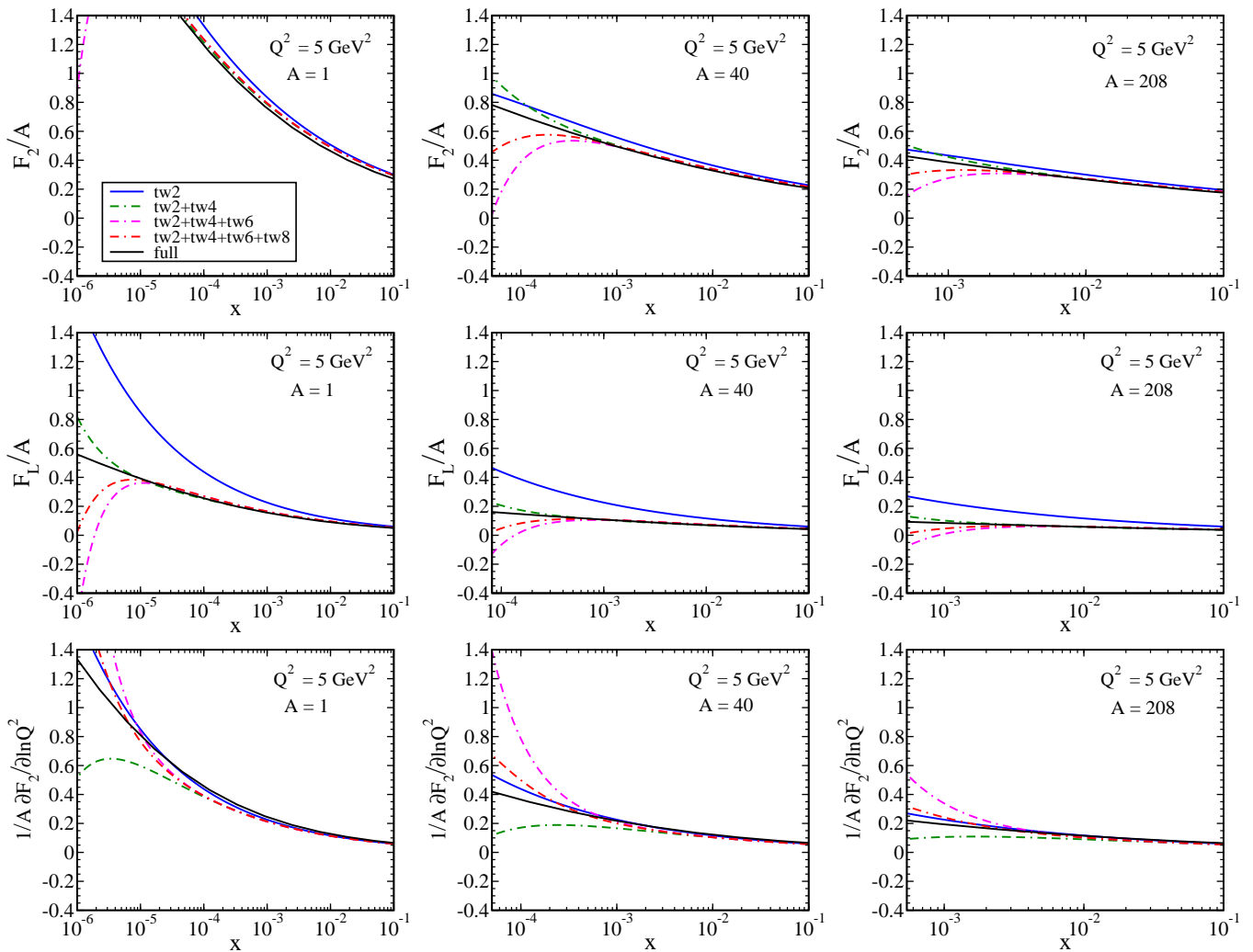

    \centering
    \includegraphics[width=.325\textwidth]{F2_Q2_5_A_1.eps}
    \includegraphics[width=.325\textwidth]{F2_Q2_5_A_40.eps}
    \includegraphics[width=.325\textwidth]{F2_Q2_5_A_208.eps} \\
    \includegraphics[width=.325\textwidth]{Fl_Q2_5_A_1.eps}
    \includegraphics[width=.325\textwidth]{Fl_Q2_5_A_40.eps}
    \includegraphics[width=.325\textwidth]{Fl_Q2_5_A_208.eps} \\ 
    \includegraphics[width=.325\textwidth]{diff_F2_Q2_5_A_1.eps}
    \includegraphics[width=.325\textwidth]{diff_F2_Q2_5_A_40.eps}
    \includegraphics[width=.325\textwidth]{diff_F2_Q2_5_A_208.eps} 
    \caption{Predictions of the distinct twists for $F_2^A(x,Q^2)$, $F_L^A(x,Q^2)$ and $\partial F_2^A /\partial ln Q^2$ considering distinct values of $A$ and assuming $Q^2 = 5$ GeV$^2$. { The minimum values for the Bjorken - $x$ variable in the plots are obtained from the condition $\xi(x,Q^2,A) = 1$ in order to guarantee that the results are presented for the region of validity of the twist expansion.   }}
    \label{Fig:difAs}
\end{figure}

\section{Results}
\label{Sec:res}

In Fig. \ref{Fig:Full_A} we present our results for the 
$x$ and atomic number dependencies of $F_2^A(x,Q^2)$, $F_L^A(x,Q^2)$ and $\partial F_2^A /\partial ln Q^2$, normalized by $A$, predicted by the phenomenological model used in our analysis assuming  $Q^2 = 2.5$ GeV$^2$ and distinct values of $A$. { The predictions will be obtained summing all twist corrections, as given by Eqs. (\ref{Eq:fs}) and (\ref{Eq:dA})}. One has that the increasing of the observables for small - $x$ is smoother when the atomic number is increased, which is directly associated with the increasing of the saturation scale with $A$. In particular, for $x \approx 10^{-5}$, the observables for $A = 208$ are suppressed by a factor of $\approx 2$ in comparison with the predictions for the proton, in agreement with the results obtained in Ref. \cite{Goncalves:2018pyn}. 
The studies performed in Refs. \cite{GayDucati:1999xg,Goncalves:2000ex} using the Glauber - Mueller approach \cite{glauber,gribov,mueller} for the the dipole - nucleus cross section have indicated that the logarithmic $Q^2$ slope of the nuclear structure function is sensitive to the saturation effects and can be used to discriminate between the linear and non - linear descriptions of the QCD dynamics. In order to verify how robust these conclusions are, in what follows we will   perform a similar analysis  to that performed in Ref. \cite{Goncalves:2000ex} using the dipole - nucleus cross section given by Eq. (\ref{Eq:dA}). Such analysis is also strongly motivated by the fact that the  high luminosities of future $eA$ colliders \cite{AbdulKhalek:2021gbh,LHeC:2020van}  will allow us to obtain a reasonable amount of data for the slope. 
Following Ref. \cite{GBW}, we will calculate the $Q^2$ and $x$ dependencies of   $\partial F_2^A /\partial ln Q^2$ for a fixed value of the photon - nucleus center - of - mass energy $W$, which is possible using that $x = Q^2 / W^2$.
In Figs. \ref{Fig:slope1} (a) and (b) we present our predictions for the $Q^2$ dependence of the slope, normalized by $A$, for different values of the atomic number  considering $W = 100$ GeV and $W = 1000$ GeV, respectively. In agreement with the results derived in Ref. \cite{Goncalves:2000ex}, we predict the presence of a maximum in the slope, with the position being dependent on the atomic number. Such result is expected when non - linear effects are present, since one of the main implications of the saturation effects are the distinct  behaviors of the gluon distribution, which is probed by the slope,  for $Q^2 > Q_s^2$ and $Q^2 < Q_s^2$, with the transition between regimes occuring for $Q^2 = Q_s^2$.  As $Q_s$ depends on $A$, the  transition is expected to occur at different points when the atomic number is modified. Moreover, the transition is also expected to be dependent on $W$, with a larger normalization due to the increasing of the gluon distribution, which is observed by comparing the results presented in the panels (a) and (b). 
 Another important aspect is that the position occurs for larger values of $Q^2$ with the increasing of the atomic number. Such dependencies are not present when the non - linear effects are disregarded, as demonstrated in Fig. \ref{Fig:slope1} (right panel), where the slope is estimated assuming that ${d^2\sigma}/{d^2\rb} = 2 (r^2 Q_{s,A}^2(x) /4) S(\rb)$, which is the linear limit of Eq. (\ref{Eq:dA}). In this case, the position of the maximum is independent of $A$ and occurs for low $Q^2$. Therefore, our results indicate that a future analysis of the slope for distinct nuclei can be useful to demonstrate the presence of the non - linear effects in the QCD dynamics, in agreement with the main conclusion from Ref. \cite{Goncalves:2000ex}.

\begin{figure}[t]
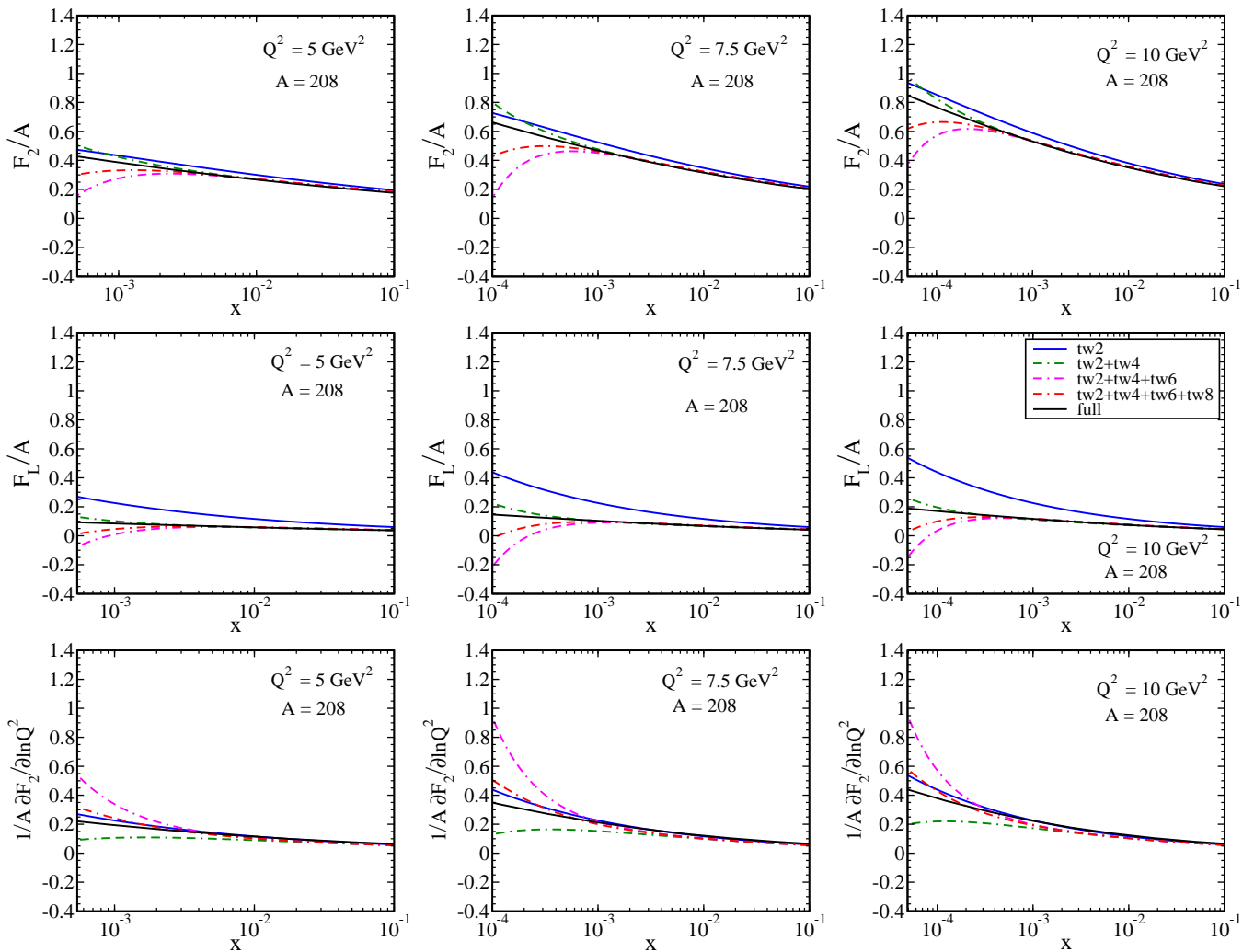

    \centering
    \includegraphics[width=0.325\textwidth]{F2_Q2_5_A_208.eps} 
    \includegraphics[width=0.325\textwidth]{F2_Q2_7_5_A_208.eps}
    \includegraphics[width=0.325\textwidth]{F2_Q2_10_A_208.eps}
\\
    \includegraphics[width=0.325\textwidth]{Fl_Q2_5_A_208.eps} 
    \includegraphics[width=0.325\textwidth]{Fl_Q2_7_5_A_208.eps}
    \includegraphics[width=0.325\textwidth]{Fl_Q2_10_A_208.eps}
\\
    \includegraphics[width=0.325\textwidth]{diff_F2_Q2_5_A_208.eps}
    \includegraphics[width=0.325\textwidth]{diff_F2_Q2_7_5_A_208.eps}
    \includegraphics[width=0.325\textwidth]{diff_F2_Q2_10_A_208.eps}

    \caption{Predictions of the distinct twists for $F_2^A(x,Q^2)$, $F_L^A(x,Q^2)$ and $\partial F_2^A /\partial ln Q^2$ assuming $A = 208$ and considering distinct values of $Q^2$. { The minimum values for the Bjorken - $x$ variable in the plots are obtained from the condition $\xi(x,Q^2,A) = 1$ in order to guarantee that the results are presented for the region of validity of the twist expansion.   }}
    \label{Fig:difQ2s}
\end{figure}

Let's now analyze the contribution of the distinct twists for $F_2^A(x,Q^2)$, $F_L^A(x,Q^2)$ and $\partial F_2^A /\partial ln Q^2$ for distinct nuclei and different values of $Q^2$. As emphasized in the previous Section, the twist expansion is valid for $\xi \lesssim 1$. As $\xi = \xi(x,Q^2,A)$, the range  in $x$ in which the predictions derived using the twist expansion are valid  is distinct for different values of $Q^2$ and $A$. In order to present our predictions only for the range of validity of Eqs. (\ref{Eq:tw2}) -- (\ref{Eq:tw8}), in what follows the plots will have distinct minimum values for the Bjorken - $x$ variable. We will focus on the atomic number and photon virtuality dependencies, which can be studied in the future $eA$ colliders, and for comparison we also will present the predictions derived using Eq. (\ref{Eq:dA}), which resums all twists (denoted full hereafter). In Fig. \ref{Fig:difAs} we present our results considering different values of $A$ and $Q^2 = 5$ GeV$^2$.  In agreement with the results derived in Refs. \cite{Bartels:2000hv,Bartels:2009tu} for $A = 1$, the impact of the higher-twist corrections on $F_2$ is small since the longitudinal and transverse twist-4 contributions have opposite signs and are almost of the same order of magnitude. { Such a conclusion is also valid for larger values of $A$, with the main difference that the impact of the higher - twist corrections becomes non-negligible  for larger values of $x$.  In contrast, the longitudinal structure function is strongly affected by the twist-4 term, which give a  sizeable negative correction to the leading-twist contribution. } The difference between the twist - 2 and full predictions for $F_L^{A = 1}$ becomes larger than 10$\%$ for $x  \lesssim 10^{-3}$ and for larger values of $x$ when the atomic number is increased. { Our results also indicate that the twist - 6 terms reduce the magnitude of $F_2^A(x,Q^2)$ and $F_L^A(x,Q^2)$ for small - $x$, while the twist - 8 contributions are positive.} For the slope, one has that higher - twist corrections become sizeable for $x \approx 10^{-3}$, but similarly to what occur in the $F_2$ case, the longitudinal and transverse contributions have opposite signal, which implies a small impact on the observable. However, it is important to emphasize that the difference between the twist - 2 and full predictions for $\partial F_2^A /\partial ln Q^2$ is larger than that predicted for $F_2^A$ in the kinematical range considered.

In Fig. \ref{Fig:difQ2s} we present the predictions of the different twists for $F_2^A(x,Q^2)$, $F_L^A(x,Q^2)$ and $\partial F_2^A /\partial ln Q^2$ assuming $A = 208$ and different values of $Q^2$. The increasing of the photon virtuality for a fixed value of $A$ implies that the validity range in $x$  of the twist expansion increases. Our results indicate that the differences between the twist - 2 and full predictions for  $F_2^A(x,Q^2)$ and $\partial F_2^A /\partial ln Q^2$ are small. In contrast, $F_L$ is strongly modified by the higher - twist corrections for all values of $Q^2$ considered in our analysis.  These results indicate that a future analysis of this observable will be useful to discriminate between linear and non - linear predictions, in agreement with  Refs. \cite{Bartels:2000hv,Goncalves:2004yc}.

A final comment is in order. As the twist expansion is valid for $\xi \lesssim 1$, the comparisons performed in Figs. \ref{Fig:difAs} and \ref{Fig:difQ2s} are restricted to the linear regime and near to the transition line. In the saturation regime, where $Q_s^2 \ge Q^2$, we expect larger differences between the twist - 2 and full predictions for a heavier nuclei, where the non - linear effects are amplified and modify the observables in a larger range of $Q^2$ and $x$. Fortunately, such a regime will also be explored in the future $eA$ colliders, in particular, at the LHeC, which will study the DIS on nuclei in a larger ($Q^2,\,x$) phase space. { It is important to emphasize that, in principle, the region where $\xi > 1$ can be explored by performing an expansion of $F_T$ and $F_L$ in positive powers of $1/\xi$, as pointed out in Ref. \cite{Bartels:2000hv}. We plan to perform such analysis in a forthcoming study.}

\section{Summary}
\label{sec:conc}
One of the main goals of  current and future colliders is the improvement of our understanding of the hadronic structure at high energies. In particular,  the search for non - linear effects in the QCD dynamics is one of the major motivations for the construction of the EIC in the US, as well as for the proposal of the LHeC and FCC - $eh$. These colliders are expected to allow for the investigation of the hadronic structure with an unprecedented precision. Electron-nucleus collisions are considered ideal to probe the nonlinear regime, since the larger parton densities
in the nuclear case, with respect to the proton case, enhance by a factor $\propto A^{1/3}$ the nuclear saturation scale, which determines the onset of non - linear effects in QCD dynamics. Motivated by this aspect and by the fact that the non - linear approaches resums higher - twist contributions, in this paper we have estimated the distinct twist terms and analyzed the impact of the different twists on $F_2^A(x,Q^2)$, $F_L^A(x,Q^2)$ and $\partial F_2^A /\partial ln Q^2$ considering different values of the Bjorken - $x$ variable, photon virtuality $Q^2$ and atomic number $A$.
We have estimated the logarithmic $Q^2$ slope for different values of the photon - nucleus center - of - mass energy. Our results indicate that the behaviour of  $F_L^A(x,Q^2)$ is strongly modified by the higher twist corrections. Furthermore, one has that a future experimental analysis of $F_L^A(x,Q^2)$ and $\partial F_2^A /\partial ln Q^2$  will allow us to probe the presence of the non - linear effects in the QCD dynamics. { The results presented in this paper motivate us to investigate the impact of the higher twist corrections on the nuclear diffractive structure functions. As observed in Refs. \cite{Motyka:2012ty,Bendova:2020hkp}, such observables are strongly sensitive to the non - linear effects.}

\begin{acknowledgments}
This work was partially supported by CNPq, CAPES, FAPERGS and INCT-FNA (Process No. 464898/2014-5) and CAPES.  V.P.G. was also partially supported by the CAS President's International Fellowship Initiative (Grant No.  2021VMA0019). 
\end{acknowledgments}



\end{document}